\documentclass[12pt, letterpaper, preprint, comicneue]{aastex63}
\pdfoutput=1
\usepackage[T1]{fontenc}
\usepackage{fontawesome}
\usepackage{color}
\usepackage{amsmath}
\usepackage{natbib}
\usepackage{ctable}
\usepackage{bm}
\usepackage[normalem]{ulem} 
\usepackage{xspace}
\usepackage{paralist}
\usepackage{fontawesome}
\usepackage{multirow}

\let\oldbibliography\thebibliography % killin' me.
\renewcommand{\thebibliography}[1]{%
  \oldbibliography{#1}%
  \setlength{\itemsep}{0pt}%
  \setlength{\parsep}{0pt}%
  \setlength{\parskip}{0pt}%
  \setlength{\bibsep}{0ex}
  \raggedright
}
\newcommand{\given}{\,|\,}

\newcommand{\bfi}[1]{\textbf{\textit{#1}}}

\newcommand{\eg}{\emph{e.g.}}
\newcommand{\ie}{\emph{i.e.}}
\newcommand{\haloflow}{{\sc HaloFlow}}

\defcitealias{simbig_letter}{H22a}
\defcitealias{simbig_mock}{H23}

\let\oldAA\AA
\renewcommand{\AA}{\text{\normalfont\oldAA}}
\newcommand{\btheta}{\boldsymbol{\theta}}
\newcommand{\bphi}{\boldsymbol{\phi}}

\newcommand{\bitem}{\begin{itemize}}
\newcommand{\eitem}{\end{itemize}}
\newcommand{\beq}{\begin{equation}}
\newcommand{\eeq}{\end{equation}}

\definecolor{orange}{rgb}{1,0.5,0}

\begin{document} \sloppy\sloppypar\frenchspacing 

\title{{\sc HaloFlow} I: Neural Inference of Halo Mass from Galaxy Photometry and Morphology}
\newcounter{affilcounter}
\author[0000-0003-1197-0902]{ChangHoon Hahn}
\altaffiliation{changhoon.hahn@princeton.edu}
\affil{Department of Astrophysical Sciences, Princeton University, Princeton NJ 08544, USA} 

\author[0000-0003-4758-4501]{Connor Bottrell}
\affil{International Centre for Radio Astronomy Research, University of Western Australia, \\ 35 Stirling Hwy, Crawley, WA 6009, Australia}
\affiliation{Kavli IPMU (WPI), UTIAS, The University of Tokyo, Kashiwa, Chiba 277-8583, Japan}
\affiliation{Center for Data-Driven Discovery, Kavli IPMU (WPI), UTIAS, The University of Tokyo,\\ Kashiwa, Chiba 277-8583, Japan}

\author[0000-0001-9299-5719]{Khee-Gan Lee}
\affiliation{Kavli IPMU (WPI), UTIAS, The University of Tokyo, Kashiwa, Chiba 277-8583, Japan}
\affiliation{Center for Data-Driven Discovery, Kavli IPMU (WPI), UTIAS, The University of Tokyo,\\ Kashiwa, Chiba 277-8583, Japan}

\begin{abstract}
We present {\sc HaloFlow}, a new machine learning approach for inferring the
mass of host dark matter halos, $M_h$, from the photometry and morphology of galaxies. 
{\sc HaloFlow} uses simulation-based inference with normalizing flows
to conduct rigorous Bayesian inference. 
It is trained on state-of-the-art synthetic galaxy images from
\cite{2023arXiv230814793B} that are constructed from the IllustrisTNG hydrodynamic 
simulation and include realistic effects of the Hyper Suprime-Cam Subaru Strategy Program 
(HSC-SSP) observations.
We design {\sc HaloFlow} to infer $M_h$ and stellar mass, $M_*$, using
$grizy$ band magnitudes, morphological properties quantifying characteristic size, concentration, and asymmetry, total measured satellite luminosity, and number 
of satellites. 
We demonstrate that {\sc HaloFlow} infers accurate and unbiased posteriors of
$M_h$.
Furthermore, we quantify the full information content in the photometric 
observations of galaxies in constraining $M_h$.
With magnitudes alone, we infer $M_h$ with $\sigma_{\log M_h} \sim 0.115$ and 0.182 dex 
for field and group galaxies. 
Including morphological properties significantly improves the precision of 
$M_h$ constraints, as does total satellite luminosity: $\sigma_{\log M_h} \sim 0.095$ and 0.132 dex.
Compared to the standard approach using the stellar-to-halo 
mass relation, we improve $M_h$ constraints by $\sim$40\%.
In subsequent papers, we will validate and calibrate {\sc HaloFlow} with 
galaxy-galaxy lensing measurements on real observational data.
\end{abstract} 
\keywords{large-scale structure of the Universe --- galaxy clusters --- galaxy groups --- Machine learning}

\section{Introduction} \label{sec:intro} 
Inferring the masses of host dark matter halos of galaxies has significant implications
for cosmology and galaxy formation. 
The abundance of most massive halos that host galaxy clusters, for instance, is sensitive 
to both the expansion history of the Universe and the growth rate of 
structure~\citep[\eg][]{voit2005, allen2011, kravtsov2012, weinberg2013, mantz2015, dodelson2016}.
It was identified as one of the most promising dark energy probes by the Dark Energy 
Task Force~\citep{detf2006}. 
With upcoming wide-field surveys such as the Vera C. Rubin Observatory Legacy Survey of Space 
and Time~\citep[LSST;][]{ivezic2019}, galaxy cluster studies are expected to significantly 
improve current constraints on dark energy.

Dark matter halos also profoundly influence the evolution and properties of galaxies 
that they host~\citep[see][for a recent review]{wechsler2018}. 
Galaxy properties, such as color, stellar mass, star formation rates, and morphology, 
have long been shown to depend significantly on local environment, which is primarily 
defined by the halo~\citep[\eg][]{oemler1974, davis1976, dressler1980, hogg2004, kauffmann2004, blanton2005, baldry2006, blanton2009, hahn2015}. 
Observational studies of this galaxy-halo connection have now firmly established that 
halo mass plays the most dominant 
role~\citep[\eg][]{tinker2011_group, moster2018_emerge, behroozi2019_um}. 

Halos can also be used to investigate the cosmic baryon distribution. 
They harbor a significant fraction of cosmic baryons, both in their stellar and interstellar media (ISM) components as well as in their extended circum-galactic media (CGM). 
The CGM, notably, exists in a warm ionized state that is not easily amenable to direct observations. 
This leads to large uncertainties regarding their overall contribution to the cosmic baryon budget, \ie~part of the so-called ``missing baryon problem''~\citep[\eg][]{fukugita2004, cen2006, bregman2007}. 
Fast radio bursts (FRBs, see \citealt{cordes2019} for a review) are a new probe that can constrain the integrated free electron column density along each line-of-sight through their observed frequency dispersion \citep[e.g.,][]{mcquinn2014,macquart2020}. 
By combining localized FRBs with detailed observations of intervening foreground galaxies, ongoing observations promise to constrain the cosmic baryon partition between the CGM and the intergalactic medium (IGM)~\citep{lee2022_frb,lee2023_frb}. 
However, since the amount and extent of CGM gas is expected to scale with the underlying halo mass \citep{xyz2019, khrykin2023cosmic},
uncertainties in the halo mass of the intervening galaxies constitute a major source of uncertainty in efforts to study the cosmic baryon distribution.

Despite the pivotal role that halos play across cosmology and galaxy evolution, 
inferring the properties of halos remains a major observational challenge.
A number of different methods are currently used to infer halo mass. 
For example, the gravitational potential of halos can be directly probed using
gravitational lensing~\citep[\eg][]{mandelbaum2006, cacciato2009, cacciato2013, mandelbaum2016_lensing, huang2020_wl}. 
Lensing mass measurements, however, require deep and high resolution imaging
especially for lower mass halos.
This makes it difficult to extend the approach to large galaxy samples. 
Satellite kinematics has also been used to infer halo mass~\citep[\eg][]{norberg2008_satkin, more2009_satkin, more2011_satkin, lange2019_satkin}. 
These studies, however, rely on the assumption of virial equilibrium,
velocity bias between the distribution of matter and satellite galaxies, 
and accurate identification of satellite galaxies.
Galaxy-based halo mass estimation methods that use the phase space
or richness information of galaxies, more broadly, have been shown to
be significantly susceptible to systematics~\citep[see][for an overview]{old2014, old2015, old2018, wojtak2018}.

There are also more indirect methods for inferring halo mass. 
Abundance matching methods assume a monotonic relation between halo mass
and galaxy stellar mass or luminosity. 
Halo masses are assigned to galaxies by matching the cumulative 
number densities of halos and galaxies~\citep{kravtsov2004, tasitsiomi2004,  vale2004, hearin2013}. 
Such approach has also been used in conjunction with halo-based group 
finding algorithms: \eg~\cite{yang2009_group, tinker2011_group, tinker2022_groupfinder}. 
These methods ultimately rely on the well-studied stellar-to-halo-mass 
relation~\citep[SHMR; see][and references therein]{wechsler2018}. 
Consequently, they do not exploit additional galaxy properties beyond
stellar mass: \eg~color, morphology. 

In this work, we present {\sc HaloFlow}, a new machine learning (ML) 
based approach that utilize the full photometric and morphological information of 
galaxies for inferring their host halo masses.  
{\sc HaloFlow} goes beyond previous ML-based halo mass estimation 
methods~\citep[\eg][]{ntampaka2015_ml, ntampaka2016_ml, calderon2019_ml, villanueva2022_camels}
in two key ways. 
First, it uses simulation-based inference (SBI) based on neural density 
estimation to conduct rigorous Bayesian inference. 
We produce full posterior distributions of halo masses that accurately quantify
the statistical uncertainties. 
Second, {\sc HaloFlow} is designed to be applied directly to observations. 
To do this, we use state-of-the-art synthetic galaxy images made with a dust radiative transfer forward model \citep{2023arXiv230814793B}. 
The images are constructed from the IllustrisTNG cosmological magneto-hydrodynamical 
simulations~\citep[hereafter ``TNG''][]{weinberger2018, pillepich2018, nelson2018} and include
the full observational realism of Subaru Hyper Suprime-Cam (HSC)
imaging data obtained through the HSC Subaru Strategy Program~\citep[HSC-SSP;][]{aihara2022}. 
This work is the first of a series of paper, where we present 
{\sc HaloFlow} and validate its performance and accuracy. 
Furthermore, we quantify the information content of photometric and 
morphological properties of galaxies for constraining halo mass. 

We begin in Section~\ref{sec:data} with a brief explanation of our 
forward-modeled synthetic images. 
We present {\sc HaloFlow} in Section~\ref{sec:haloflow}. 
Afterwards, we present the results of applying {\sc HaloFlow} in 
Section~\ref{sec:results} and discuss their implications in Section~\ref{sec:discuss}. 
\section{Data} \label{sec:data}
One of the main components of SBI is a forward-model of the observable. 
In this work, we use forward-modeled images and corresponding 
photometric and morphological measurements of galaxies from TNG.
Below, we briefly describe the forward-model. 
We refer readers to \cite{2023arXiv230814793B} for full details.

\subsection{TNG simulations} \label{sec:tng}
The TNG simulations\footnote{\url{https://tng-project.org}} are a suite 
of publicly available cosmological magneto-hydrodynamical 
simulations~\citep{weinberger2018, pillepich2018, 2018MNRAS.475..648P, springel2018, 2018MNRAS.480.5113M, 2018MNRAS.477.1206N, nelson2018, 2019ComAC...6....2N} 
that use 
{\sc Arepo}\footnote{\url{https://arepo-code.org}}~\citep{springel2010}
to track the 
co-evolution of gas, stars, dark matter, super-massive black holes,
and magnetic fields from $z = 127$ to $z=0$. 
The model includes subgrid treatments for the formation and evolution of 
stellar populations, black hole growth, radiative cooling, stellar and black
hole feedback, and magnetic fields.
TNG includes simulations run at three sets of volumes and resolutions. 
This work makes use of data derived from the highest-resolution TNG50 simulation, which is run in a $(35\,{\rm cMpc}/h)^3$ box with baryonic mass resolution of $M_b\approx8.5 \times 10^4 \mathrm{M}_{\odot}$. 
TNG50 galaxies with stellar masses of $M_* \ge 10^9 M_\odot$ (i.e.\ $\gtrsim 10^4$ star particles) have been shown to be reasonably resolved with robust stellar structures \citep{2021MNRAS.508.5114L,2023MNRAS.525.5614L}.

\subsection{Synthetic images and measurements}

TNG50 galaxies spanning $0.\leq z \leq0.7$ and $M_* \ge 10^9 M_\odot$ were forward-modeled into synthetic images from the HSC-SSP by \cite{2023arXiv230814793B}. The forward-modeling procedure first uses the $\mathtt{SKIRT}$ dust radiative transfer code\footnote{\url{https://skirt.ugent.be}} \citep{baes2011,camps2015,camps2020} to make noise/background-free, high-resolution (idealized) images in the HSC $grizy$ bands\footnote{\url{https://www.tng-project.org/explore/gallery/bottrell23i}} \citep{kawanomoto2018} and several supplementary filters spanning $0.3-5$ microns. The radiative transfer model uses the \cite{bruzual2003} stellar population synthesis (SPS) library to model light from stellar populations older than 10 Myr and assumes a \cite{chabrier2003} initial mass function. Continuum and nebular line emission from young stellar populations embedded in birth clouds (star particles ages $<10$ Myr) are modeled with the MAPPINGS III library \citep{groves2008}. 
Dust is not explicitly modelled in TNG, therefore, we carry out post-processing to model the absorption/scattering of light by dust. 
We ascribe dust densities to gas cells using the method described by \cite{popping2022}, in which the dust-to-gas mass ratio scales with gas metallicity \citep{2014A&A...563A..31R}. The dust model further assumes a \cite{weingartner2001} Milky Way dust grain composition and size distribution. Each galaxy is `observed' along four different orientations in order to increase the overall statistical sample.

\begin{figure*}
    \includegraphics[width=\textwidth]{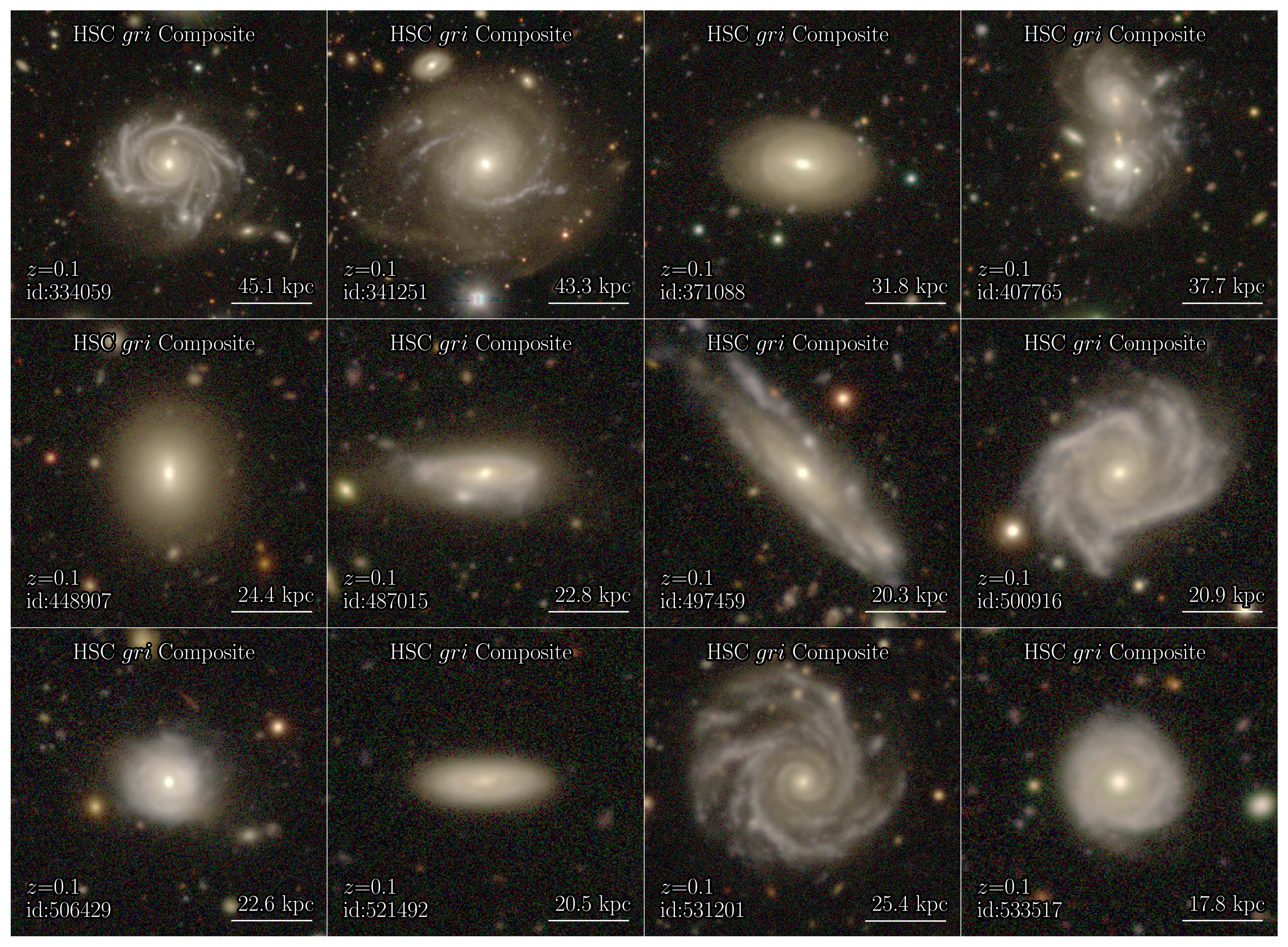}
    \caption{\label{fig:sims}
        Synthetic forward-modeled $gri$ composite images of galaxies 
        tailored to the HSC-SSP observations. 
        The images are constructed from TNG50 using $\mathtt{SKIRT}$ and 
        stellar population synthesis.
        They include the realistic survey effects as found in the 
        HSC-SSP. 
        We show 12 randomly selected central galaxies from our data set. 
        The magnitudes, $R_{\rm eff}$, and $CAS$ morphology parameters used in this work are 
        measured from these forward-modeled images. 
    }
\end{figure*}

Next, the survey effects of the HSC-SSP Public Data Release 3~\citep[PDR3;][]{aihara2022} are applied to the output images from $\mathtt{SKIRT}$. An adapted version of $\mathtt{RealSim}$\footnote{\url{https://github.com/cbottrell/RealSim}}~\citep{bottrell2019} is used to assign insertion locations, perform flux calibration, spatially rebin to the HSC pixel scale, convolve the images with reconstructed HSC point-spread functions, and insert into realistic HSC-like images\footnote{\url{https://www.tng-project.org/explore/gallery/bottrell23}}. 
The images have sufficiently large fields-of-view that they include extended structure, satellites, and nearby group/cluster members.
Eisert et al (in prep) make a detailed comparison of these TNG50 mocks to real HSC galaxy images and show that their morphologies broadly agree. 

After the TNG50 mock images were generated, we carried out photometric and morphological measurements using the GALIGHT \footnote{\url{https://github.com/dartoon/galight}} surface-brightness decomposition software \citep{2020ApJ...888...37D}. 
Specifically, we measure magnitudes and effective radii, $R_{\rm eff}$, from parametric S\'{e}rsic fits and morphologies quantified by the Concentration, Asymmetry, and Smoothness/Clumpiness ({\em CAS}) parameters~\citep{abraham1994, abraham1996, bershady2000, conselice2003}. Since these quantities are measured from the synthetic HSC images, they include realistic measurement uncertainties. In Figure~\ref{fig:sims}, we show forward-modeled $gri$ composite images of 12 randomly selected central galaxies from the data set. 

For this paper, we focus on central galaxies at $z = 0.1$, classified using the TNG Friends-of-Friends \citep[FoF;][]{davis1985} group catalog. 
For each central galaxy, we compile its true stellar mass ($M_*$), true host halo mass ($M_h$),  
and the `observed' $griyz$-band S\'{e}rsic magnitudes $\bfi{X}_{\rm mag}$, and morphological properties 
$\bfi{X}_{\rm morph} = \{R_{{\rm eff}, X}, c_{X}, A_{X}\}$. The variables $R_{{\rm eff}, X}, c_{X}, A_{X}$ correspond to the characteristic size, concentration, and asymmetry in the $X = g, r, i, z, y$ bands. 
We also compile the total measured luminosity, $L_{{\rm sat.},X}$, and number, $N_{\rm sat}$, of satellites brighter than  $M_r < -18$ within each individual group. 
In total, we use 7,468 photometric measurements from 1,867 simulated central galaxies. We reserve a subset of 125 random central galaxies with $M_* > 10^{9.5}M_\odot$ for testing {\sc HaloFlow} and use the rest for training. 

\section{{\sc haloFlow}} \label{sec:haloflow}
To infer the posterior, $p(\btheta\given\bfi{x})$, of $\btheta = \{M_*, M_h\}$ 
given observational measurements, $\bfi{x}$, we use the \haloflow~SBI framework. 
SBI\footnote{also known as ``likelihood-free'' or ``implicit-likelihood'' inference}
enables inference using only a generative model of mock observations. 
While various SBI approaches have been applied to inference problems in
astronomy~\citep[\eg][]{cameron2012, weyant2013, hahn2017b, alsing2018, wong2020, zhang2021}, 
we specifically use an approach based on neural density estimation. 
In particular, we use ``normalizing flow'' models~\citep{tabak2010, tabak2013}, 
following the SBI approach in {\sc SEDFlow}~\citep{hahn2022a}.

Flows use neural networks to learn an extremely flexible and bijective transformation, 
$f: x \mapsto z$, that maps a complex target distribution to a simple base distribution, 
$\pi(\bfi{z})$.
The target distribution, in our case, is the posterior $p(\btheta\given\bfi{x})$ and 
$f$ is designed to be invertible and have a tractable Jacobian. 
This is so that the posterior can be evaluated from $\pi(\bfi{z})$ by change of 
variables. 
We choose a multivariate Gaussian for $\pi(\bfi{z})$, which makes the posterior easy 
to sample and evaluate.
Out of the different flow architectures, we use Masked Autoregressive
Flow~\citep[MAF;][]{papamakarios2017} models as implemented in the $\mathtt{sbi}$ Python package~\citep{greenberg2019, tejero-cantero2020}.

We train a flow, $q_{\bphi}$, with hyperparameters, $\bphi$, to best 
approximate the posterior, 
$q_{\bphi}(\btheta\given\bfi{x}) \approx p(\btheta\given\bfi{x})$.
We split the simulated galaxies into a training and validation set with 90/10 split. 
Then for galaxies $\{(\btheta_i, \bfi{x}_i)\}$ in the training set, we maximize the 
total log-likelihood $\sum_i \log q_{\bphi}(\btheta_i\given \bfi{x}_i)$.
This is equivalent to minimizing the KL divergence between 
$p(\btheta, \bfi{x}) = p(\btheta\given\bfi{x}) p(\bfi{x})$ and
$q_{\bphi}(\btheta\given\bfi{x}) p(\bfi{x})$. 
We use the {\sc Adam} optimizer~\citep{kingma2017} with a learning rate of 
$5\times10^{-4}$. 
To prevent overfitting, we stop training when the log-likelihood evaluated 
on the validation set fails to increase after 20 consecutive epochs. 

To determine our final normalizing flow, we train a large number ($\sim$1000) of flows 
with architectures determined using the {\sc Optuna} hyperparameter optimization 
framework~\citep{akiba2019optuna}.
We then select five flows, $q_{\bphi}^j$, with the lowest validation losses and construct 
our final flow by ensembling them: 
$q_{\bphi}(\btheta\given\bfi{x})  =  \sum_{j=1}^5 q_{\bphi}^j(\btheta\given\bfi{x})/5$. 
Ensembling flows with different initializations and architectures improves the 
accuracy of our normalizing flow~\citep{lakshminarayanan2016, alsing2019}.

$q_{\bphi}$ implicitly includes a prior, $p(\btheta)$, which is set by the $M_*$ and 
$M_h$ distribution of our training data set. 
Without any corrections, this prior reflects the stellar and halo mass functions.
Posteriors with this prior would favor low $M_*$ and $M_h$ values since there
are more galaxies with lower $M_*$ and $M_h$.
We correct for this implicit prior and impose uniform priors on $\log M_*$ and $\log M_h$ using 
the \cite{handley2019maxent} maximum entropy prior method. 
In practice, for a sample drawn from our posterior, 
$\btheta' \sim q_{\bphi}(\btheta\given\bfi{x})$, we impose an importance weight of
$1/p(\btheta')$.
This ensures that we assume uniform priors on $M_*$ and $M_h$. 

In this work, we make use of different sets of photometric measurements: photometry, 
photometry and morphology, and etc.
For each set of observables, we repeat the entire procedure above and train a separate 
ensembled flow.

\begin{figure*}
    \includegraphics[height=0.45\textwidth]{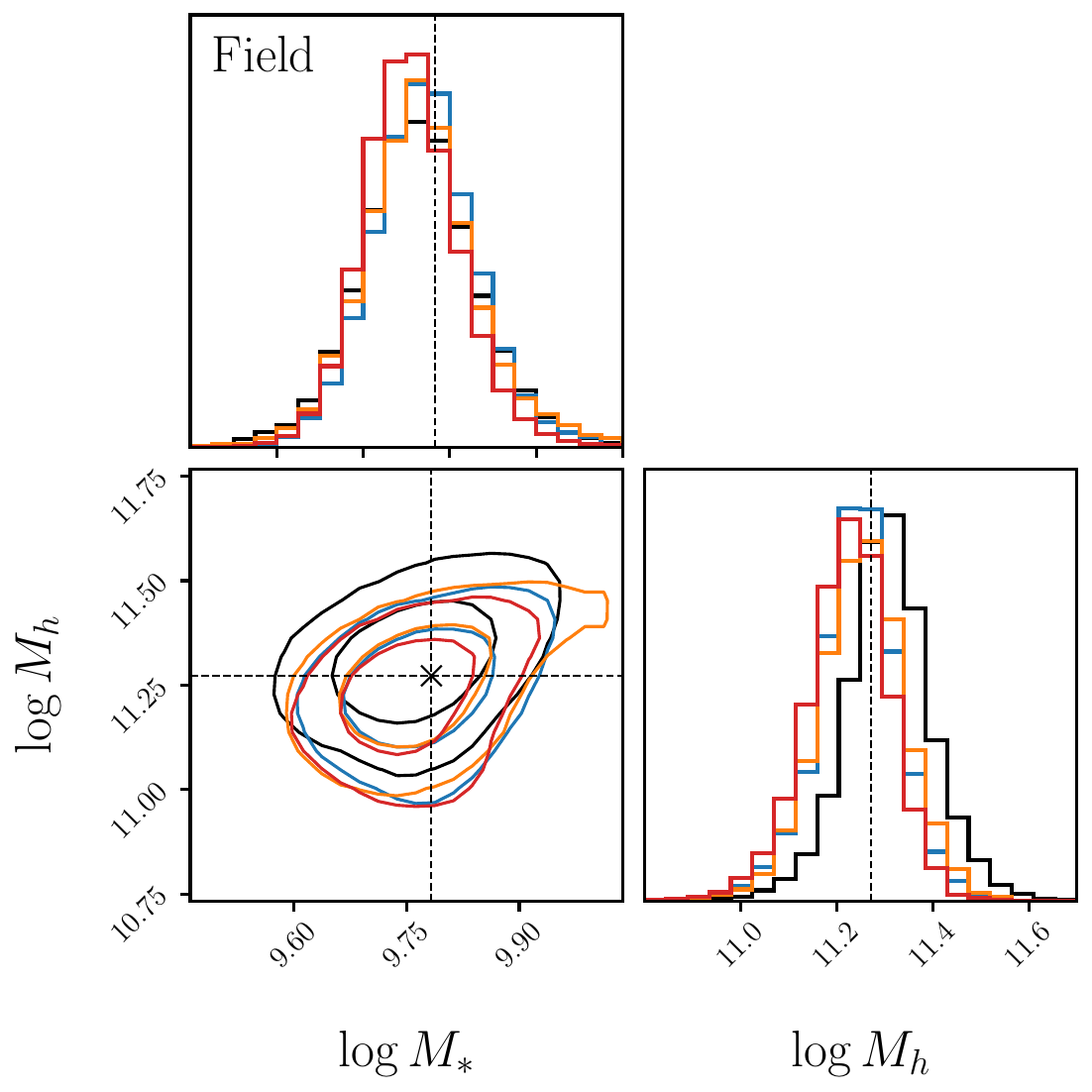}
    \includegraphics[height=0.45\textwidth]{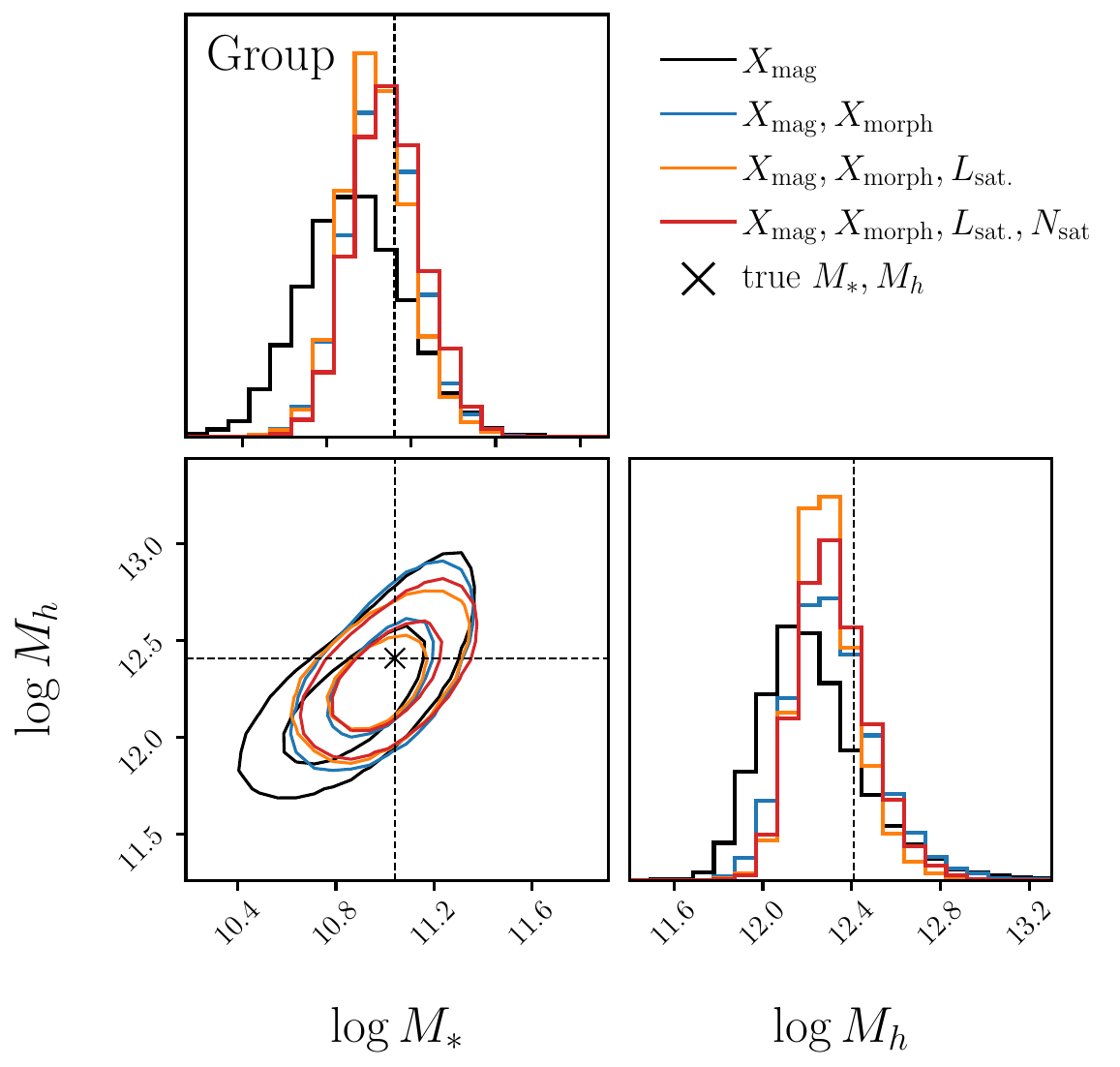}
    \caption{\label{fig:posterior}
    $M_*$ and $M_h$ posteriors inferred from different sets of photometric measurements: 
    $\{X_{\rm mag}\}$ (black), 
    $\{X_{\rm mag}, X_{\rm morph}\}$ (blue), 
    $\{X_{\rm mag}, X_{\rm morph}, L_{\rm sat}\}$ (orange), and
    $\{X_{\rm mag}, X_{\rm morph}, L_{\rm sat}, N_{\rm sat}\}$ (red). 
    The contours mark the $68^{\mathrm{th}}$ and $95^{\mathrm{th}}$  percentiles.
    The left and right set of panels show the posterior for a field and group galaxy, 
    respectively.
    We also mark the true $M_*$ and $M_h$ value of the arbitrarily selected simulated galaxy (black $\times$). 
    All of the {\sc HaloFlow} posteriors are consistent with the true
    $M_*$ and $M_h$ values. 
    Furthermore, the posteriors demonstrate that morphology, satellite 
    luminosity, and richness contribute significant additional 
    constraining power for $M_*$ and $M_h$. 
    }
\end{figure*}

\begin{table}
    \caption{{\sc HaloFlow} median posterior standard error (dex) in stellar mass, $\sigma_{\log M_*}$,  and halo mass, $\sigma_{\log M_h}$, predictions for field and group galaxies relative to ground truth.}
    \centering
    \begin{tabular}{|c|c|c|c|c|}
        \hline
        \multicolumn{1}{|c|}{{\sc HaloFlow} Input} &
        \multicolumn{2}{c|}{Field Galaxies} & 
        \multicolumn{2}{c|}{Group Centrals} \\
        \hfill & $\sigma_{\log M_*}$ & $\sigma_{\log M_h}$ & $\sigma_{\log M_*}$ & $\sigma_{\log M_h}$ \\ [3pt]
        \hline
        $\{X_{\rm mag}\}$ & 0.096 & 0.115 & 0.151 & 0.182 \\
        $\{X_{\rm mag}, X_{\rm morph}\}$ & 0.078 & 0.105 & 0.109 & 0.149 \\
        $\{X_{\rm mag}, X_{\rm morph}, L_{\rm sat}\}$ & 0.078 & 0.095 & 0.118 & 0.138 \\
        $\{X_{\rm mag}, X_{\rm morph}, L_{\rm sat}, N_{\rm sat}\}$ & 0.073 & 0.095 & 0.108 & 0.132\\[3pt] \hline
        Standard SHMR Method &  & 0.175 &  &  0.208 \\
        \hline
    \end{tabular}
    \label{tab:posteriors}
\end{table}

\section{Results} \label{sec:results}
In Figure~\ref{fig:posterior}, we present the posteriors of $M_*$ and $M_h$ 
for an arbitrarily selected field galaxy (left) and group galaxy (right) inferred using {\sc HaloFlow} with different sets of observables: 
$\{X_{\rm mag}\}$ (black), 
$\{X_{\rm mag}, X_{\rm morph}\}$ (blue), 
$\{X_{\rm mag}, X_{\rm morph}, L_{\rm sat}\}$ (orange), and
$\{X_{\rm mag}, X_{\rm morph}, L_{\rm sat}, N_{\rm sat}\}$ (red). 
We mark the $68^{\mathrm{th}}$ and $95^{\mathrm{th}}$ percentiles of the posteriors as contours as well as the
true $M_*$ and $M_h$ values of the galaxy (black x). The field galaxy resides in a $M_h =  10^{11.27}M_\odot$ 
halo and has no satellite galaxies brighter than $M_r < -18$; the group galaxy resides in a 
$M_h = 10^{12.41}M_\odot$ halo and has 7 satellites brighter than $M_r < -18$. 

All of the {\sc HaloFlow} posteriors are consistent with each other and in excellent
agreement with the true $M_*$ and $M_h$.
For the field galaxy, there is a significant improvement from including $X_{\rm morph}$.
However, there is expectedly little improvement from including $L_{\rm sat}$ or 
$N_{\rm sat}$ since it has no satellite galaxies.
The central column of Table \ref{tab:posteriors} summarizes the median standard deviation 
of {\sc HaloFlow} posteriors all field central galaxies in the test sample. 
With the inclusion of $X_{\rm morph}, L_{\rm sat}, N_{\rm sat}$, we improve the precision of $M_*$ and $M_h$ 
constraints for the field galaxy sample by $\sim$0.023 and 0.020 dex --- a $\sim$20\% improvement. 

Meanwhile, for the group central galaxy in Figure \ref{fig:posterior}, including 
each additional photometric measurement significantly improves the precision of the 
$M_*$ and $M_h$ constraints. 
The median standard deviation of {\sc HaloFlow} posteriors for all group centrals in the test sample are shown in 
the right column of Table~\ref{tab:posteriors} for each set of observables. 
Photometric measurements beyond magnitudes, improve the precision of $M_*$ and $M_h$ 
constraints by $\sim$0.043 and 0.050 dex --- a $\sim$30\% improvement.

For both field and group centrals, the {\sc HaloFlow} posteriors
firmly demonstrate that galaxy morphology encodes significant 
information on both $M_h$ and $M_*$. 
This confirms previous works that found connections between 
morphology and local environment~\citep[\eg][]{dressler1980, wilman2012, perez-millan2023}.
Our results also show that the photometric measurements of satellite galaxies 
are informative of $M_h$ and $M_*$. 
This is consistent with \cite{tinker2021_satlum}, who found that total 
satellite luminosity is an excellent proxy for $M_h$.
Beyond confirming previous works, with {\sc HaloFlow} we precisely quantify
the information content of these observables for constraining  $M_*$ and $M_h$. 
Furthermore, {\sc HaloFlow} provides a rigorous Bayesian inference framework for 
actually leveraging these photometric measurements.

\begin{figure}
    \begin{center}
        \includegraphics[width=0.8\textwidth]{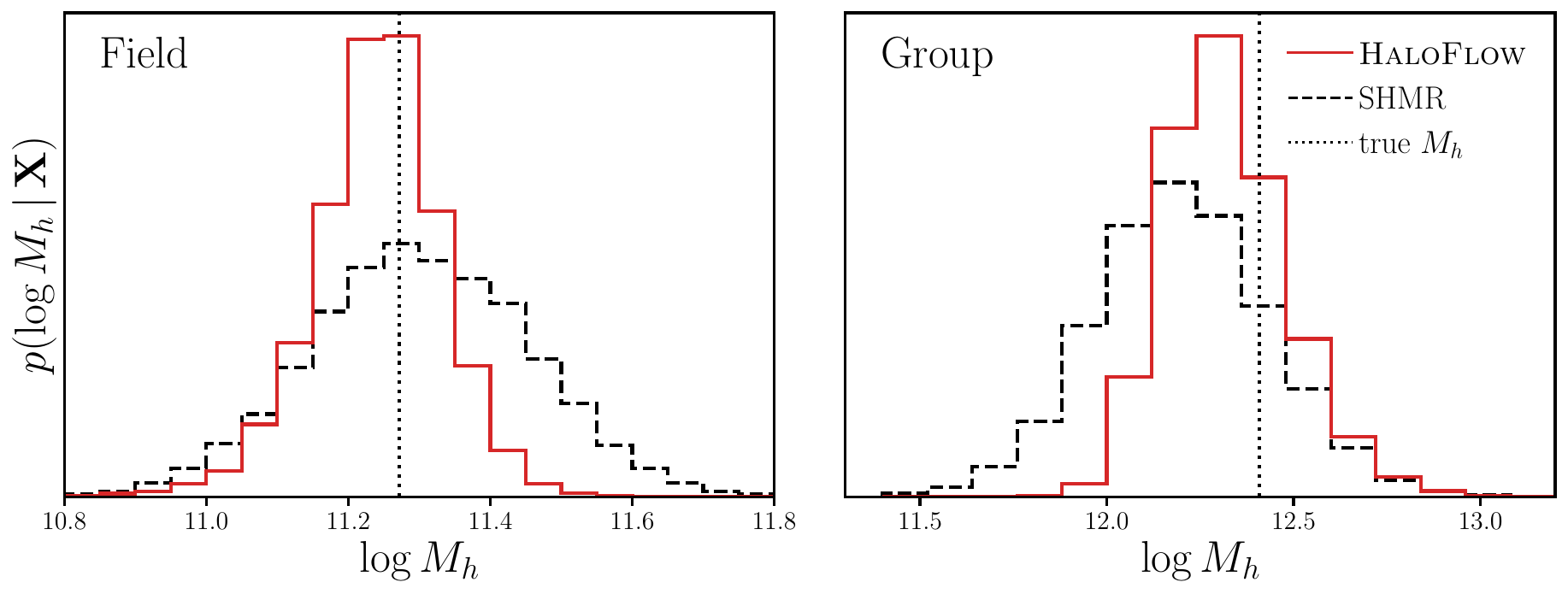}
        \caption{\label{fig:straw}
        {\sc HaloFlow} $M_h$ posterior using 
        $\{X_{\rm mag}, X_{\rm morph}, L_{\rm sat}, N_{\rm sat}\}$ (red) compared to 
        the $M_h$ constraints from the standard approach using the SHMR 
        (black dashed) for the same field (left) and group galaxy (right) as in Figure~\ref{fig:posterior}.
        The $M_h$ constraint for the standard method is derived from measuring $M_*$ 
        from photometry using SED modeling and then translating it to $M_h$ using 
        the SHMR.
        With {\sc HaloFlow}, we can exploit the constraining power of $X_{\rm morph}$, 
        $L_{\rm sat}$, and $N_{\rm sat}$ and improve $M_h$ constraints by
        $\sim$0.08 dex ($\sim$40\%) over the standard approach. 
        }
    \end{center}
\end{figure}

We further compare the {\sc HaloFlow} $M_h$ posterior (red) to constraints derived from
the standard approach (black dashed) for a field (left) and group galaxy (right) in 
Figure~\ref{fig:straw}.
We use the same galaxies as in Figure~\ref{fig:posterior}. 
For the standard approach, we derive the $M_h$ constraint by first measuring $M_*$ from 
photometry using SED modeling and then converting $M_*$ to $M_h$ using the SHMR. 
We first draw samples from the $M_*$ posterior, $M_*' \sim p(M_*\given X_{\rm mag})$, 
then sample $M_h' \sim p(M_h \given M_*')$ given by the SHMR. 
We estimate $p(M_*\given X_{\rm mag})$ using {\sc HaloFlow} and estimate $p(M_h \given M_*')$ 
from the TNG simulations\footnote{
We derive the mean SMHR directly from TNG50 (Section~\ref{sec:tng}) and use $\sigma_{\log M_h}\sim 0.15$ dex based on \cite{wechsler2018}, due to the limited number of galaxies in TNG50.}. 
This ensures that we do not introduce any biases from discrepant assumptions in the SED modeling 
(\eg~stellar library, initial mass function, dust modeling) and provides the most ``apples to apples'' comparisons with {\sc HaloFlow}.
For the {\sc HaloFlow} posterior, we use the posterior from 
$\{X_{\rm mag}, X_{\rm morph}, L_{\rm sat}, N_{\rm sat}\}$.
With the standard approach, we derive $\sigma_{\log M_h} = 0.175$ and 
$0.208$ dex for satellite and group central galaxies, respectively
(Table~\ref{tab:posteriors}). 
Our {\sc HaloFlow} $M_h$ constraints are significantly, $\gtrsim 0.080$ 
and $0.076$ dex, tighter than the standard approach. 
This corresponds to $\sim 40\%$ tighter constraints on $M_h$.

In addition to the tighter $M_h$ constraints, {\sc HaloFlow} provides a fully consistent 
framework for deriving $M_h$ directly from observations. 
In our comparison, as mentioned above, we implement an idealized version of the standard 
approach where the SED modeling and the SHMR are consistent 
by construction. 
However, in practice, the $M_*$ derived from SED modeling are not consistent with 
the $M_*$ values used in the SHMR from simulations. 
The $M_*$ from SED modeling is a measurement that depends on the specific assumptions 
of stellar population synthesis.  
Depending on modeling choices, the inferred $M_*$ can vary by 
$\sim$ 0.1 dex~\citep{pacific2023}. 
Observational effects, such as backgound subtraction~\citep{bernardi2017}, can also
significantly impact the $M_*$ inferred from SED modeling. 
Meanwhile, the $M_*$ in SHMR is a theoretical quantity, typically derived from summing
up the masses of all the star particles in a subhalo. 
Any discrepancies in the $M_*$ measured from SED modeling versus the simulation will bias 
the inferred $M_h$. 
In {\sc HaloFlow}, all of these effects are consistently accounted in the forward model. 

\begin{figure*}
\begin{center}
    \includegraphics[width=0.8\textwidth]{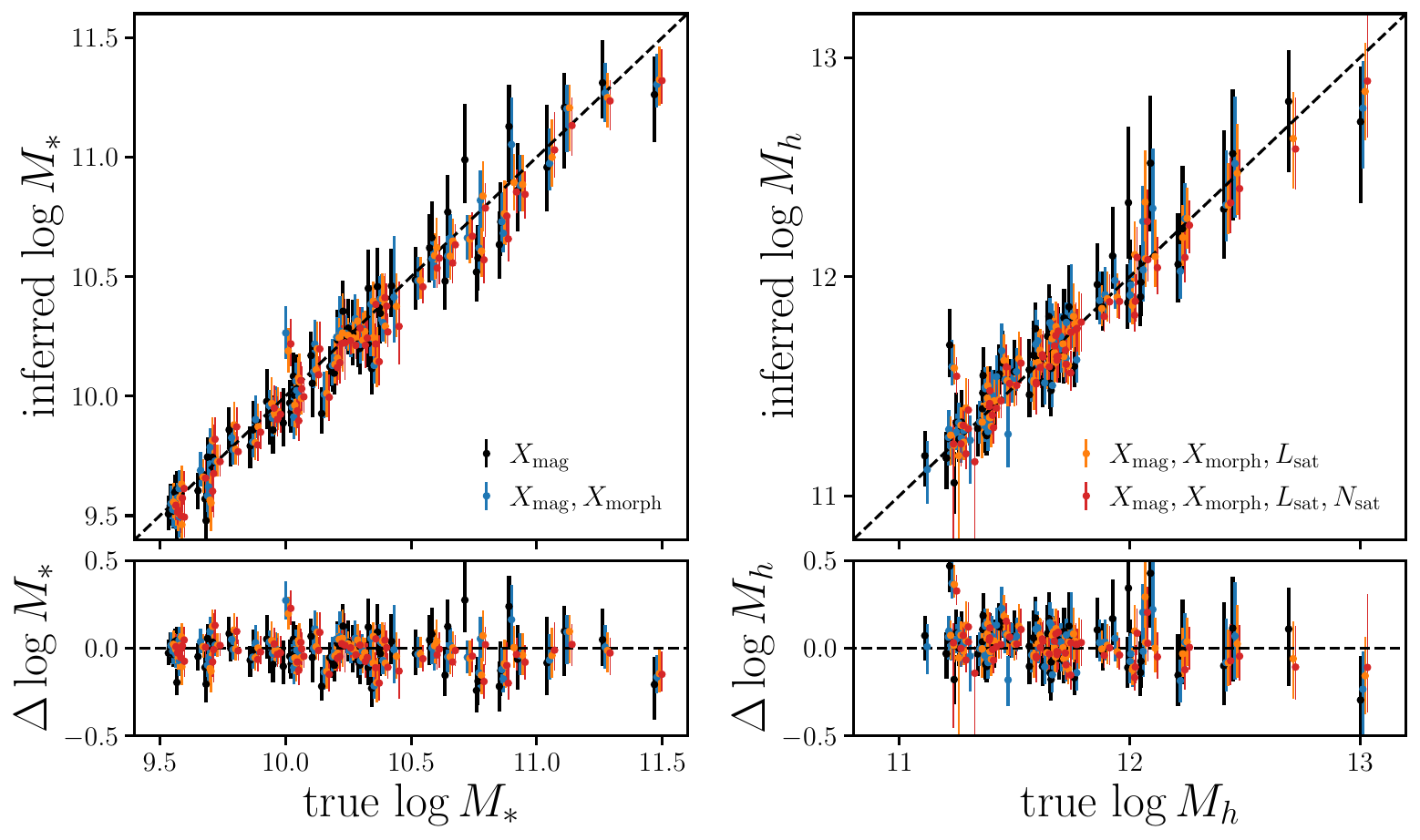}
    \caption{\label{fig:valid_errorbar}
    Inferred {\sc HaloFlow} $M_*$ and $M_h$ versus the true values for 
    galaxies in our test sample (top panels).
    The bottom panels show the residuals. 
    We include the posteriors from the different sets of photometric 
    measurements (black, blue, orange, red) and represent their $68^{\mathrm{th}}$ percentiles with the error bars. 
    We include only a subset of the test sample for clarity. 
    Overall, {\sc HaloFlow} accurately infers the true $M_*$ and $M_h$. 
    The comparison further confirms that $X_{\rm morph}$, $L_{\rm sat}$, 
    and $N_{\rm sat}$ each significantly tighten the $M_*$ and $M_h$ 
    posteriors for certain galaxies. 
    }
\end{center}
\end{figure*}

Next, we validate the posteriors from {\sc HaloFlow}. 
In Figure~\ref{fig:valid_errorbar}, we compare the {\sc HaloFlow} $M_*$ and $M_h$ inferred from 
$\{X_{\rm mag}\}$ (black), 
$\{X_{\rm mag}, X_{\rm morph}\}$ (blue), 
$\{X_{\rm mag}, X_{\rm morph}, L_{\rm sat}\}$ (orange), and 
$\{X_{\rm mag}, X_{\rm morph}, L_{\rm sat}, N_{\rm sat}\}$ (red) to the true values (top panels). 
We show the residuals in the bottom panels. 
The error bars represent the $68^{\mathrm{th}}$ percentiles of the {\sc HaloFlow} posteriors. 
For clarity, we only include 60 of 125 galaxies from the test sample in Figure~\ref{fig:valid_errorbar}. 
The comparison illustrates that overall {\sc HaloFlow} accurately infers the true $M_*$ and $M_h$. 
Furthermore, it shows that incorporating $X_{\rm morph}$, $L_{\rm sat}$,  and $N_{\rm sat}$ 
significantly tightens the $M_*$ and $M_h$ posteriors. 

\begin{figure*}
\begin{center}
    \includegraphics[width=0.495\textwidth]{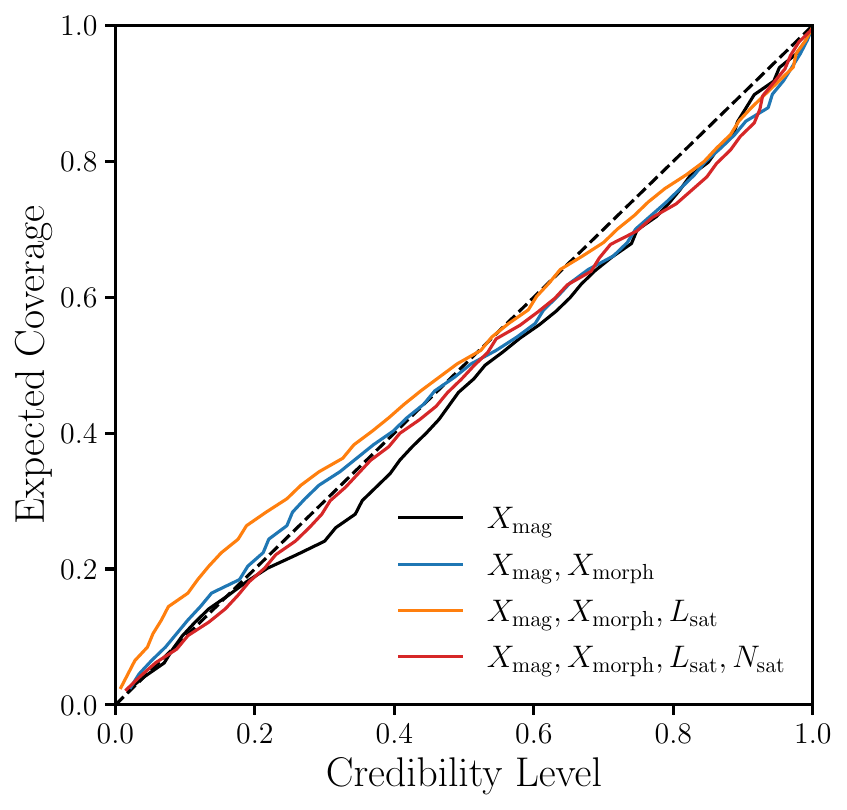}
    \caption{\label{fig:valid_tarp}
    Coverage test validating the accuracy of our {\sc HaloFlow} posterior 
    estimate using $\{X_{\rm mag}\}$ (black), $\{X_{\rm mag}, X_{\rm morph}\}$ (blue), 
    $\{X_{\rm mag}, X_{\rm morph}, L_{\rm sat}\}$ (orange), and 
    $\{X_{\rm mag}, X_{\rm morph}, L_{\rm sat}, N_{\rm sat}\}$ (red). 
    The test is calculated using the test sample. 
    The black-dashed line represents an optimal estimate of the true posterior. 
    {\sc HaloFlow} provides a near optimal estimate of the true posterior for 
    all sets of photometric measurements.}
\end{center}
\end{figure*}

As additional validation, we use the \cite{lemos2023} ``data to random point'' (DRP) 
coverage test. 
For each test sample, we evaluate distances between samples drawn from {\sc HaloFlow} 
posteriors and a random point in parameter space. 
We compare these distances to the distance between the true $M_*$, $M_h$ and the
random point to derive an estimate of the expected coverage probability. 
This approach is necessary and sufficient to show that a posterior estimator is 
optimal.
In Figure~\ref{fig:valid_tarp}, we present the DRP coverage test of the {\sc HaloFlow} 
posteriors for each of the photometric measurements. 
Significant discrepancies from the true posterior (black-dashed) can reveal 
whether the posterior estimates are underconfident, overconfident, or biased. 
In our case, we find no significant discrepancies.
Hence, {\sc HaloFlow} provides near optimal estimates of the true 
posteriors for all of the adopted combinations of observables.

\section{Discussion} \label{sec:discuss}
{\sc HaloFlow} leverages state-of-the-art forward modeled galaxy images to exploit
additional photometric information that significantly improves constraints on 
$M_*$ and $M_h$. 
Hence, a primary determining factor for the fidelity of {\sc HaloFlow} is the 
quality of the forward-model. 
Below, we discuss the caveats and limitations of our forward-model. 

First, our forward model is based on a particular galaxy formation model: that 
adopted in the TNG cosmological magneto-hydrodynamical simulation. 
However, previous works have revealed significant discrepancies among the properties 
of galaxy populations predicted by different state-of-the-art galaxy formation models. 
\cite{hahn2019}, for instance, found significant discrepancies among the $M_*$-star 
formation rate relations of Illustris, EAGLE, and MUFASA hydrodynamical simulations.
Nevertheless, a number of works have demonstrated the success of TNG at reproducing
a wide range of observations: 
\emph{e.g.} galaxy color bimodality~\citep{nelson2018},  sizes~\citep{genel2018}, optical morphologies \citep{2019MNRAS.483.4140R,2021MNRAS.501.4359Z},
mass-metallicity relation~\citep{torrey2019}, and low redshift quasar luminosity 
function and black hole mass - stellar bulge mass relation~\citep{weinberger2018}. 
Furthermore, one of the main relations that {\sc HaloFlow} exploits from TNG is the SHMR. 
The SHMR in galaxy formation models is typically calibrated against constraints from 
observations and, thus, consistent across different models~\citep{wechsler2018}. 
In TNG, the SHMR is not explicitly calibrated against observational constraints; however, 
it is indirectly calibrated since the subgrid physics is optimized to match the $z=0$ 
stellar mass function~\citep{pillepich2018}.

Our forward model also relies on a specific mass-to-light conversion framework. 
$\mathtt{SKIRT}$ is responsible for generating our synthetic observables from the 
galaxy physical properties predicted by TNG. 
The transfer model for the HSC-SSP synthetic images uses the MAPPINGS III SED library % \citep{groves2008} 
to model emission from stellar populations forming in birth-clouds and \cite{bruzual2003} 
templates for older stellar populations, assuming a \cite{chabrier2003} initial mass function. 
These libraries are standard choices in the literature for both SED modeling (\eg~MAGPHYS~\citealt{dacunha2015magphys}, BEAGLE~\citealt{chevallard2016beagle}, BAGPIPES~\citealt{carnall2018bagpipes}) and for radiative transfer modeling (\eg~\citealt{2015MNRAS.447.2753T,2017MNRAS.470..771T,2023MNRAS.518.5522C,2023MNRAS.519.4920G}). 
However, for inferring galaxy properties (\emph{i.e.} the inverse problem of forward modeling), 
different SED modeling choices can produce significant 
discrepancies in galaxy properties, \eg~$M_*$ and SFR~\citep{pacific2023}. 

The second component of the mass-to-light conversion framework is the dust model, which handles the scattering and absorption of stellar light by dust. 
The synthetic images from \cite{2023arXiv230814793B} use a dust model calibrated to gas mass, gas metallicity, and dust mass estimates for galaxies in the local Universe \citep{2014A&A...563A..31R,popping2022}.
While empirical, the model is physically intuitive --- the gas-to-dust mass ratio should scale with the metal content of the gas at temperatures where dust grains can form. However, there is considerable scatter about the empirical relation for observed galaxies. 
An in-depth exploration of different SED/dust modeling choices is beyond the scope of this work. 
But given that significant differences in the inferred physical properties of galaxies arise 
from the choice of SED and dust modeling this is an area that warrants investigation.

In future works, we will investigate the choices made in our forward-model. 
For instance, we will train {\sc HaloFlow} using additional simulations based on 
other galaxy formation models~(\eg~SIMBA,~\citealt{dave2019}; EAGLE,~\citealt{crain2015, schaye2015}).
We will also utilize alternative SPS and dust models.
With different forward-models we can improve the robustness of {\sc HaloFlow} by 
ensuring that it does not learn relationships among galaxy and halo properties 
specific to a single galaxy formation model. 
We will also be able to extensively cross-validate {\sc HaloFlow} and confirm the 
robustness of the inferred host halo properties. 

These additional simulations will also improve the accuracy of {\sc HaloFlow}, 
especially at high $M_*$ and $M_h$ where there are currently a limited number 
of simulated galaxies. 
A larger set of simulated galaxies will also enable more systematic exploration
of additional photometric observables that can inform $M_*$ and $M_h$. 
Furthermore, with {\sc HaloFlow} trained on different galaxy formation models, 
we can precisely quantify the information content of the galaxy-halo connection in
each model. 
The information content will not only serve as an informative statistic of the
models but also by comparing across the models we will be able to inform galaxy
formation. 

In subsequent papers, we will also test and calibrate the {\sc HaloFlow} 
$M_h$ constraints of observed galaxies and groups in HSC-SSP against constraints 
from  galaxy-galaxy weak lensing measurements~\citep[\eg][]{rana2022}. 
We will also compare them to dynamical mass estimates of groups identified in GAMA~\citep{driver2022}. 
Once validated, we will apply {\sc HaloFlow} to intervening halos 
observed by the FLIMFLAM spectroscopic survey~\citep{lee2022_frb} 
targeting FRB foreground fields to enhance their constraints on the CGM baryonic fraction.

Lastly, we limited the $L_{\rm sat}$ and $N_{\rm sat}$ measurements to satellite 
galaxies with $M_r < -18$ in this work. 
Including fainter galaxies in $L_{\rm sat}$ and $N_{\rm sat}$ would further improve 
the precision of {\sc HaloFlow}. 
Future observations from DESI and Rubin, which will probe significantly fainter galaxies,
will be able to take advantage of the additional gains from {\sc HaloFlow}. 

\section{Summary} \label{sec:summary}
We present {\sc HaloFlow}, a framework for inferring host halo masses from the 
photometry and morphology of galaxies using simulation-based inference (SBI) with
normalizing flows. 
{\sc HaloFlow} is specifically tailored to galaxies in the Hyper Suprime-Cam
Subaru Strategy Program (HSC-SSP) and, thus, leverages state-of-the-art 
synthetic galaxy images that model the realistic effects of the HSC-SSP 
observations~\citep{2023arXiv230814793B}. 
These images are constructed from the TNG hydrodynamic simulations using the
$\mathtt{SKIRT}$ dust radiative transfer and an adapted version of $\mathtt{RealSim}$. 

% quick summary of SBI 
We train {\sc HaloFlow} using 7,468 photometric measurements of 1,867 central 
galaxies made on the synthetic HSC images using GALIGHT. 
The measurements include $grizy$-band magnitudes ($X_{\rm mag}$), 
$CAS$ morphological parameters ($X_{\rm morph}$), 
total measured satellite luminosities ($L_{\rm sat}$), and 
number of satellites ($N_{\rm sat}$). 
{\sc HaloFlow} uses normalizing flows to perform neural density estimation  of
the posterior, $p(\btheta \given \bfi{x})$, of $\btheta = \{M_*, M_h\}$ given
the photometric measurements. 
We follow the SBI approach of \cite{hahn2022a} with two additional steps. 
First, our final flow is derived from ensembling the five flows with the 
lowest validation losses. 
Second, we correct for the implicit prior on $M_*$ and $M_h$ set by the stellar 
and halo mass functions using the \cite{handley2019maxent} maximum entropy prior 
method. 
We train separate flows, 
$q_\phi(\btheta \given \bfi{x}) \approx p(\btheta \given \bfi{x})$, 
for different sets of photometric measurements, $\bfi{x} = \{X_{\rm mag}\}$, 
$\{X_{\rm mag}, X_{\rm morph}\}$, $\{X_{\rm mag}, X_{\rm morph}, L_{\rm sat}\}$, 
$\{X_{\rm mag}, X_{\rm morph}, L_{\rm sat}, N_{\rm sat}\}$.

% summary of results
When we apply {\sc HaloFlow} to a subset of 125 random central galaxies with 
$M_* > 10^{9.5} M_\odot$ excluded from the training we find: 
\begin{itemize}
    \item {\sc HaloFlow} successfully infers posteriors that are consistent with 
    the true $M_*$ and $M_h$ for every set of photometric measurements.
    We further validate the accuracy of the posteriors using the \cite{lemos2023robust}
    DRP cover test and confirm that {\sc HaloFlow} provides near optimal estimates
    of the true posteriors. 
    \item Comparison of the {\sc HaloFlow} posteriors firmly demonstrate that
    galaxy morphology encodes significant information on both $M_*$ and $M_h$. 
    This confirms and quantifies the known connection between morphology and local 
    environment.  
    The {\sc HaloFlow} posteriors also show that satellite measurements can further
    improve $M_h$ constraints. 
    With these additional observables, we can improve $M_h$ constraints by $\sim$20 and 30\% for field and group galaxies, respectively. 
    \item  With all of our photometric measurements, we can constrain $M_*$ and $M_h$ 
    with precision levels of 
    $\sigma_{\log M_*}\sim 0.073$ and $\sigma_{\log M_h}\sim 0.095$ dex for 
    field galaxies and 
    $\sigma_{\log M_*}\sim 0.108$ and $\sigma_{\log M_h}\sim 0.132$ dex for group 
    galaxies.
    Our {\sc HaloFlow} $M_h$ constraints are $\sim$40\% tighter than standard 
    $M_h$ methods based on the SHMR.
\end{itemize}

% discussion and subsequent works
{\sc HaloFlow} uses SBI to leverage state-of-the-art synthetic galaxy images. 
Its fidelity is, therefore, determined by the quality of the forward model used to 
simulate the images. 
Our forward model relies on the TNG galaxy formation model, which has been
show to successfully reproduce a wide range of observations. 
It also relies on $\mathtt{SKIRT}$, which uses standard modeling choices in the 
literature. 
Nevertheless, in future work we will go beyond these choices and use additional
galaxy formation models and alternative SPS and dust models to improve the 
robustness of {\sc HaloFlow}.
We will also further validate {\sc HaloFlow} using alternative methods for inferring
$M_h$ based on galaxy-galaxy weak lensing. 
Afterwards, we will apply {\sc HaloFlow} to HSC-SSP and infer $M_h$ for a variety of
applications: \eg~constrainig $M_h$ of halos in the foreground of FRBs to 
place stringent constraints on the CGM baryonic fraction.

\section*{Acknowledgements}
It's a pleasure to thank Marc Huertas-Company, Mike Walmsley, Ken Wong, and John Wu 
for useful discussions.
This work was supported by the AI Accelerator program of the Schmidt Futures Foundation. CB gratefully acknowledges support from the Natural Sciences and Engineering Council of Canada and the Forrest Research Foundation.
This work was substantially performed using the Princeton Research Computing resources
at Princeton University, which is a consortium of groups led by the Princeton Institute 
for Computational Science and Engineering (PICSciE) and Office of Information Technology’s 
Research Computing.
Kavli IPMU is supported by World Premier In-
ternational Research Center Initiative (WPI), MEXT, Japan. 

This work made use of premier images captured by Subaru Telescope on the summit of Maunakea, Hawaii. We acknowledge the cultural, historical, and natural significance and reverence that Maunakea has for the indigenous Hawaiian community. We are deeply fortunate and grateful to share in the opportunity to explore the Universe from this mountain. The Hyper Suprime-Cam (HSC) collaboration includes the astronomical communities of Japan and Taiwan, and Princeton University. The HSC instrumentation and software were developed by the National Astronomical Observatory of Japan (NAOJ), the Kavli Institute for the Physics and Mathematics of the Universe (Kavli IPMU), the University of Tokyo, the High Energy Accelerator Research Organization (KEK), the Academia Sinica Institute for Astronomy and Astrophysics in Taiwan (ASIAA), and Princeton University. Funding was contributed by the FIRST program from Japanese Cabinet Office, the Ministry of Education, Culture, Sports, Science and Technology (MEXT), the Japan Society for the Promotion of Science (JSPS), Japan Science and Technology Agency (JST), the Toray Science Foundation, NAOJ, Kavli IPMU, KEK, ASIAA, and Princeton University.

\bibliography{haloflow} 
\end{document}